\documentstyle[12pt,epsf]{article}
\begin{document}
\newcommand{\be}[1]{\begin{equation} \label{(#1)}}
\newcommand{\ee}{\end{equation}}
\newcommand{\ba}[1]{\begin{eqnarray} \label{(#1)}}
\newcommand{\ea}{\end{eqnarray}}
\newcommand{\nn}{\nonumber}

\begin{center}
  {\large {\bf 
  Degenerate Neutrinos and CP Violation}}\\[1.5cm]
    {Ara N. Ioannisian$^{1,2}$ and N. A. Kazarian$^{1}$}\\[0.35cm]

{\em 1) Institute for Theoretical Physics and Modeling, Halabian 34, 
Yerevan-36, 
Armenia} \\ 
{\em 2) Yerevan Physics Institute, Alikhanian Br.\ 2, Yerevan-36, Armenia}

\end{center} 

\centerline {\bf Abstract} 

We have studied mixing and masses of three left handed Majorana neutrinos 
in the model, which assumes exactly degenerate neutrino masses at some 
"neutrino unification" scale.
Such a simple theoretical ansatz naturally leads to quasidegenerate 
neutrinos.
The neutrino mass splittings induced by renormalization effects. 
In the model we found that the parameters of the neutrino physics 
(neutrino mass spectrum, mixing angles and CP violation phases) 
are strongly intercorrelated to each other. 
From these correlations we got strong bounds on the parameters 
which could be checked in the oscillation experiments.

\newpage

In the present article  we have studied the model of degenerate in mass 
neutrinos at some high energy in the general CP violation case.

We add to the basic Lagrangean the dimension--five
operator
\begin{eqnarray}
\frac{\lambda_0 \delta_{ab} }{M_X} (\phi\ell_a)(\ell_b \phi)~+~H.~c.
\label{ans}
\end{eqnarray}
where $\ell_a$ denote the three lepton doublets and $\phi$ is the
standard Higgs doublet. 
The breaking of the
electroweak symmetry due to a non-zero vacuum expectation value (VEV)
$<\phi>$ will generate, in addition to the known SM masses, the
seesaw-type neutrino mass operator $M_\nu$
\be{mass}
M_\nu =  \frac{<\phi>^2}{M_X}  \lambda_0 \ .
\ee

In the basis in which the
charged lepton Yukawa couplings are diagonal (weak basis), $\lambda_0
\delta_{ij}$ gets transformed into
\be{l5weak}
\Lambda = \lambda_0 U^* U^\dagger
\ee
where $U^\dagger$ is  a matrix which is diagonalized 
the charged lepton Yukawa couplings. 

In that basis we get
\begin{equation}
{\cal L}\supset{\lambda_0 <\phi >^2 \over M_X} \nu^T \nu
-{g\over\sqrt2}\overline{e_L}\gamma^\mu W^-_\mu U\nu ñ+~H.~c.
\label{eqn:lagr}
\end{equation}

In general, the mixing matrix $K$
is characterized by three mixing angles and three $CP$ violating
phases, one Dirac plus two Majorana-type phases
and can be written as
\begin{eqnarray}
\label{K}
&&\hspace{-2.5cm}K \ = \ \left(\matrix{c_{12}c_{13}&s_{12}c_{13}&s_{13}e^{-i\phi}\cr
-s_{12}c_{23}-c_{12}s_{23}s_{13}e^{i\phi}&
c_{12}c_{23}-s_{12}s_{23}s_{13}e^{i\phi}&
s_{23}c_{13}\cr s_{12}s_{23}-c_{12}c_{23}s_{13}e^{i\phi}&
-c_{12}s_{23}-s_{12}c_{23}s_{13}e^{i\phi}&c_{23}c_{13}}\right)
 \nonumber \\
\cdot
&&\left(\matrix{1&0&0\cr0&e^{i\alpha_1}&0\cr0&0&e^{i\alpha_2}}
\right)
\label{matrix}
\end{eqnarray}

It is clear that with our ansatz
(\ref{ans}) we have a freedom of performing any
arbitrary real $3\times3$ rotation of the neutrino mass eigenstates 
\cite{Branco:1998bw}. 
However once the neutrino mass degeneracy is lifted by quantum corrections the
mixing effects require the full matrix $K$ with a
non-trivial relationship amongst the mixing angles.

Now we turn to the renormalization effects.  The renormalization
group equations (RGEs) for the $\Lambda$ coefficients characterizing
the dimension-5 non-renormalizable terms can easily be written down
both for the SM and MSSM.  The flavour independent
corrections to the $\Lambda = \O(1)$ coefficients are irrelevant for
our discussion. The flavour-dependent corrections are due to lepton
Yukawa couplings and, to a good approximation, are determined by the
$\tau$ Yukawa coupling.
Moreover the supersymmetry can induce flavour-dependent
threshold corrections associated with slepton mass splitting
which can dominate over the $\tau$ Yukawa corrections.

The quantum corrections to our ansatz (\ref{ans})
in the weak basis are given by 
\begin{equation}
\label{Mnu}
M_\nu \ = \  c \ m_0 ( 1 + R ) \ U^* U^\dagger ( 1 + R ))
\end{equation}
where $c$ is a common flavour-independent renormalization $\O(1)$ 
factor.  
The correction $R$ (calculated
in the electroweak charged lepton mass eigenstate basis) consists of
two parts - the renormalization group correction
and the electroweak scale threshold corrections.  
Assuming no lepton flavour violation in other sectors of the theory 
(e.g. in the case of supersymmetry) the matrix $R$ is diagonal.

Without loss of generality $R$ can be written as
\begin{eqnarray}
\label{R} 
R \ = \ r \left( \matrix { 1 & & \cr & 0 & \cr & & \epsilon } \right) \ 
\end{eqnarray}

In \cite{Chankowski:2000fp} it was shown that only when $| \epsilon | < 1 $ the ansatz 
(\ref{ans}) leads to the correct phenomenological results.

The ordinary perturbation calculus tells that the
off-diagonal entries of the matrix 
\begin{equation}
\label{A}
A \ = K^T \ R \ K^* \ + \ K^\dagger \ R \ K
\end{equation}
should be zero. We therefore require that
\begin{equation}
\label{A0}
A_{12}=A_{13}=A_{23}=0 \ .
\end{equation}

The corrected neutrino masses are given by
\begin{eqnarray}
\label{masses}
m_1 \ = m \ ( 1 \ + \ r A_{11} ) \nonumber \\
m_2 \ = m \ ( 1 \ + \ r A_{22} ) \\
m_3 \ = m \ ( 1 \ + \ r A_{33} ) \nonumber
\end{eqnarray}
where $m=c \ m_0$.

Inserting $K$ and $R$ into eq. (\ref{A}), taking into account  
eqs. (\ref{A0}) and (\ref{masses}) and assuming maximal atmospheric 
mixing angle we arrive to the following relations
\begin{equation}
\label{x}
\tan^2 \theta_{12} \ = \ - \frac{\cos (\beta-\phi) \ \cos (\alpha-\beta)}{\cos
\beta \ \cos
(\alpha+\phi-\beta)  },
\end{equation}
\begin{equation}
\label{s13}
s_{13}  =    \left| \frac{\cos \alpha \ \cos (\alpha - \beta)}{\cos
(\alpha+\phi-\beta) ( \cos (\alpha-\phi)-\tan^2 \theta_{12} \ \cos(
\alpha+\phi)  )}
\right|^{1/2},
\end{equation}   
and
\begin{equation}
\label{delta}
\frac{\Delta m_{21}^2}{\Delta m_{32}^2} \ = \ \frac{\cos 
(2 \theta_{12})}{\sin^2
\theta_{12}}+\frac{4 \ \cos \alpha \ \cos \phi}{\cos (\alpha-\phi)
-\tan^2 \theta_{12} \ \cos( \alpha+\phi)}. 
\end{equation}

On the figs. we have pointplotted the predicted areas of the model on 
neutrino oscillation parameters.

\vspace{1cm}
\noindent
\vskip 1cm 
\centerline{\epsfxsize=140mm\epsfbox{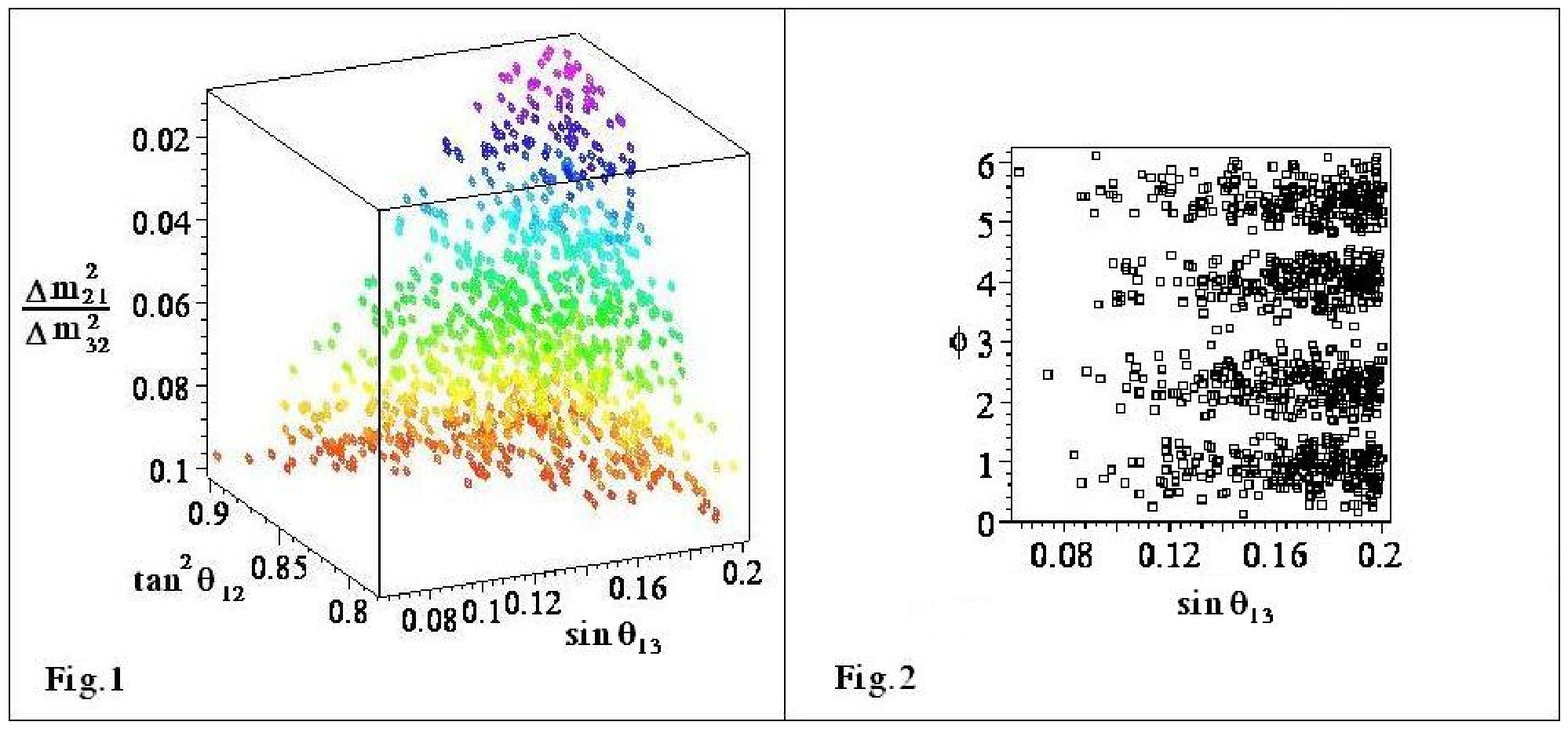}}
\vspace{5mm}
Fig. 1 - The area in the 
$\Delta m^2_{12} / \Delta m^2_{23}$ - $\tan^2\theta_{12}$ - 
$\sin \theta_{13}$ space predicted in our model. 
For a lower value of the solar mixing angle 
($\theta_{12}$) the model predicts a large ratio 
$\Delta m^2_{12} / \Delta m^2_{23}$ and a large "reactor" mixing angle 
($\theta_{13}$). 

Fig. 2 - The areas in the 
$\phi$ - $\sin \theta_{13}$ plane predicted in our model.
\noindent

\newpage 

\noindent
\vskip 1cm 
\centerline{\epsfxsize=140mm\epsfbox{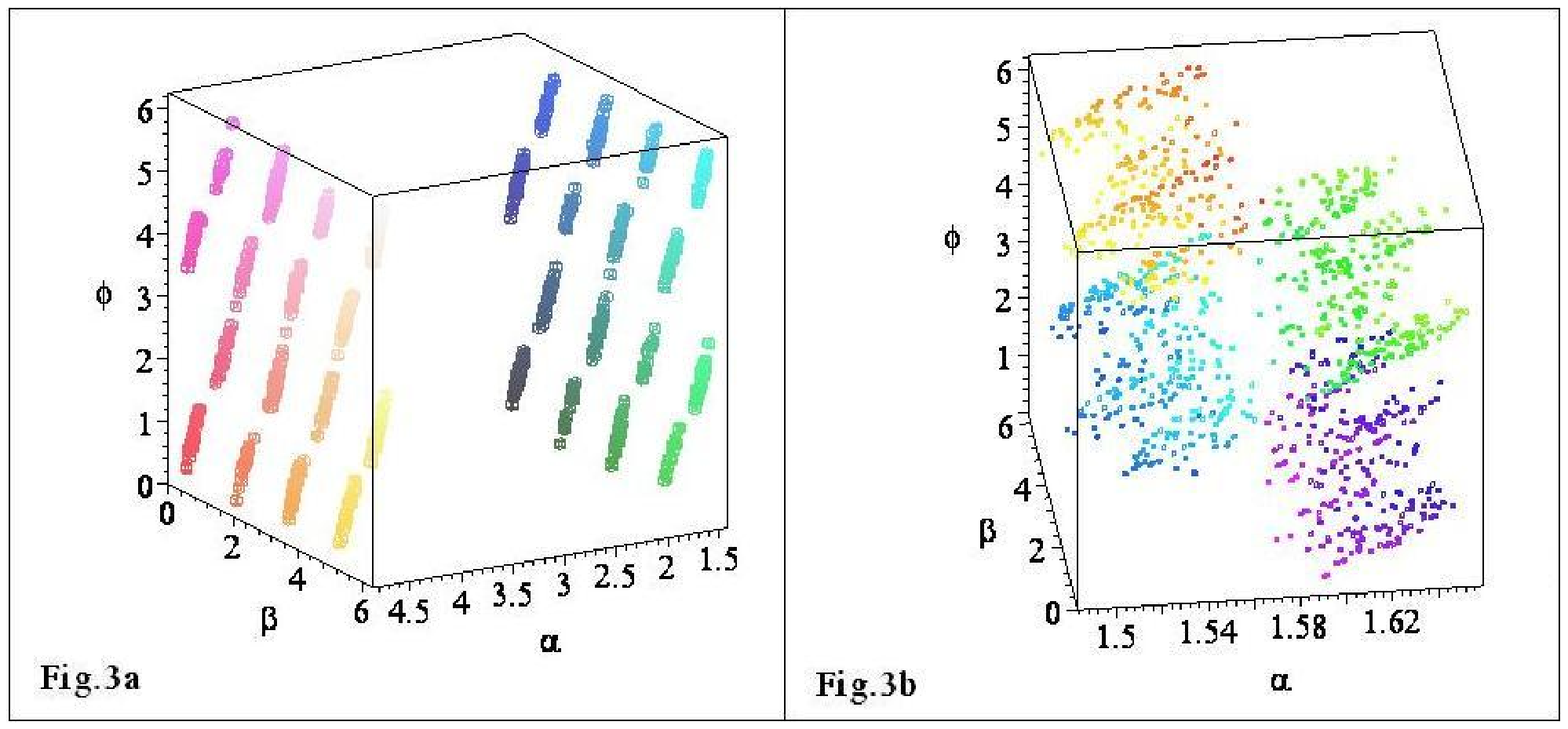}}
Fig. 3a$^1$ - The areas of the charged leptonic 
mixing matrix phases ( Majorana 
$\alpha$, $\beta$ ; Dirac $\phi$ ) predicted in our model. 
The solutions are only for $\alpha \simeq \pi / 2  + n \pi$ 
and $\phi - 2 \beta \simeq k \pi$ ( k, n are integers). 

Fig. 3b$^1$ - The areas of the charged leptonic 
mixing matrix phases ( Majorana 
$\alpha$, $\beta$ ; Dirac $\phi$ ) predicted in our model 
around value $\alpha \simeq \pi / 2$.
\noindent
\vspace{1cm}
\noindent
\vskip 1cm 
\centerline{\epsfxsize=140mm\epsfbox{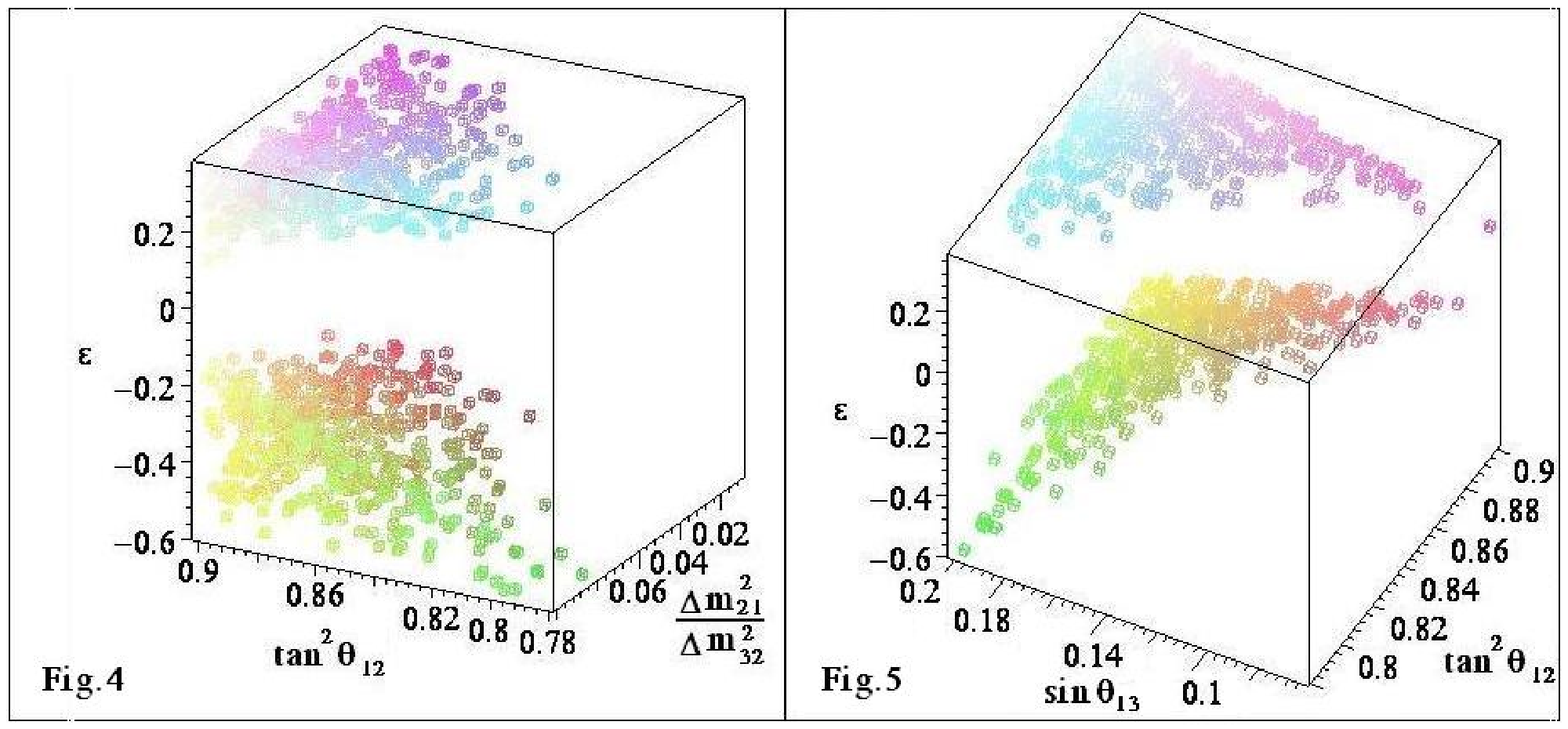}}
Fig. 4$^1$ - The parameter $\epsilon$ of our model 
depending on $\Delta m^2_{12} / \Delta m^2_{23}$ and 
$\tan^2\theta_{12}$. 

Fig. 5$^1$ - The parameter $\epsilon$ of our model 
depending on $\sin \theta_{13}$ and $\tan^2\theta_{12}$.
\noindent

{\footnote{The different colors are only used for better 3d visuality.}} 

\begin{thebibliography}{99}
\bibitem{Chankowski:2000fp}
P.~H.~Chankowski, A.~Ioannisian, S.~Pokorski and J.~W.~F.~Valle,
Phys.\ Rev.\ Lett.\  {\bf 86} (2001) 3488
[arXiv:hep-ph/0011150].
\bibitem{Branco:1998bw}
G.~C.~Branco, M.~N.~Rebelo and J.~I.~Silva-Marcos,
Phys.\ Rev.\ Lett.\  {\bf 82}, 683 (1999)
[arXiv:hep-ph/9810328].
\end{thebibliography}
\end{document}